\documentclass[sigconf,anonymous=false]{acmart}

\AtBeginDocument{%
  \providecommand\BibTeX{{%
    \normalfont B\kern-0.5em{\scshape i\kern-0.25em b}\kern-0.8em\TeX}}}

\copyrightyear{2024}
\acmYear{2024}
\setcopyright{rightsretained}
\acmConference[Websci Companion '24]{16th ACM Web Science Conference}{May 21--24, 2024}{Stuttgart, Germany}
\acmBooktitle{16th ACM Web Science Conference (Websci Companion '24), May 21--24, 2024, Stuttgart, Germany}\acmDOI{10.1145/3630744.3658415}
\acmISBN{979-8-4007-0453-6/24/05}

\setcopyright{none}
\renewcommand\footnotetextcopyrightpermission[1]{}
\setcopyright{none}

\begin{document}

\title{Investigating Bias in Political Search Query Suggestions by Relative Comparison with LLMs}

\author{Fabian Haak}
\email{fabian.haak@th-koeln.de}
\orcid{0000-0002-3392-7860}
\affiliation{
  \institution{TH K\"oln (University of Applied Sciences)}
  \city{Cologne}
  \country{Germany}}

\author{Björn Engelmann}
\orcid{0009-0000-7074-9066}
\email{bjoern.engelmann@th-koeln.de}
\affiliation{
  \institution{TH K\"oln (University of Applied Sciences)}
  \city{Cologne}
  \country{Germany}}

\author{Christin Katharina Kreutz}
\orcid{0000-0002-5075-7699}
\email{christin.kreutz@th-koeln.de}
\affiliation{
  \institution{TH K\"oln (University of Applied Sciences)}
  \city{Cologne}
  \country{Germany}}

\author{Philipp Schaer}
\orcid{0000-0002-8817-4632}
\email{philipp.schaer@th-koeln.de}
\affiliation{
  \institution{TH K\"oln (University of Applied Sciences)}
  \city{Cologne}
  \country{Germany}
}

\renewcommand{\shortauthors}{Haak et al.}

\begin{abstract}
  Search query suggestions affect users' interactions with search engines, which then influences the information they encounter.
  Thus, bias in search query suggestions can lead to exposure to biased search results and can impact opinion formation. 
  This is especially critical in the political domain.
  Detecting and quantifying bias in web search engines is difficult due to its topic dependency, complexity, and subjectivity.
  The lack of context and phrasality of query suggestions emphasizes this problem. 
  In a multi-step approach, we combine the benefits of large language models, pairwise comparison, and Elo-based scoring to identify and quantify bias in English search query suggestions.
  We apply our approach to the U.S. political news domain and compare bias in Google and Bing.
\end{abstract}



\received{29 February 2024}

\received[accepted]{1 April 2024}

\maketitle

\section{Introduction \& Related Work}\label{sec:introduction}
Search engines, notably Google and Bing, are widely trusted sources for information, particularly in areas like political news~\cite{ray2020}. 
Their impact on forming political opinions has been shown~\cite{epstein2015}. Users articulate their information needs through search queries while interacting with search query suggestions provided by the search engine to save time or when their information need is not established (i.e., their knowledge on the subject is low)~\cite{niu}.
This interaction and the subsequent results are susceptible to biases~\cite{Introna2000}.
Although (media) bias is investigated profoundly (e.g., \cite{Spinde2023MediaBias}), few works explore bias in online search beyond search results. 
Research on bias in search queries is scarce.
\citet{robertson_auditing_2018} and \citet{kulshrestha2019} investigate bias in online search, including biased queries' effects. 
Many studies focus on bias in search query suggestions as stand-ins for user-logged search queries, particularly in the political domain~\cite{Bonart_2019, haak_perception-aware_2021, HaakS22}. The investigation of bias in search query suggestions has the further benefit of identifying bias in queries that influence the information search of users with less established information needs.
A general scarcity of query datasets and the lack of context and phrasality hinder the development of effective means to identify bias in search query suggestions.
Further, the subjective nature of bias (posed by the user's perception, expectation, and knowledge) and context dependency pose challenges to the automatic quantification of bias at the level of single search queries without considering search results or additional information.
Effective means of identifying and, even more so, quantifying biases in search query suggestions would be a large step towards fairer and safer access to political information and search engines in general.

\section{Bias Identification \& Quantification}\label{sec:methodology}
We aim to overcome the problems of detecting bias in query suggestions. 
Fine-tuning and training language models to detect bias has proven to be an effective means to identify bias in the news domain; however, this requires the availability of labeled datasets~\cite{Spinde2023MediaBias}.
These datasets do not exist for query suggestions.
By utilizing \textit{gpt-4-1106-preview}, we can omit the need for fine-tuning and achieve effective results.

\begin{figure}[t]
    \centering
    \includegraphics[width=1\linewidth]{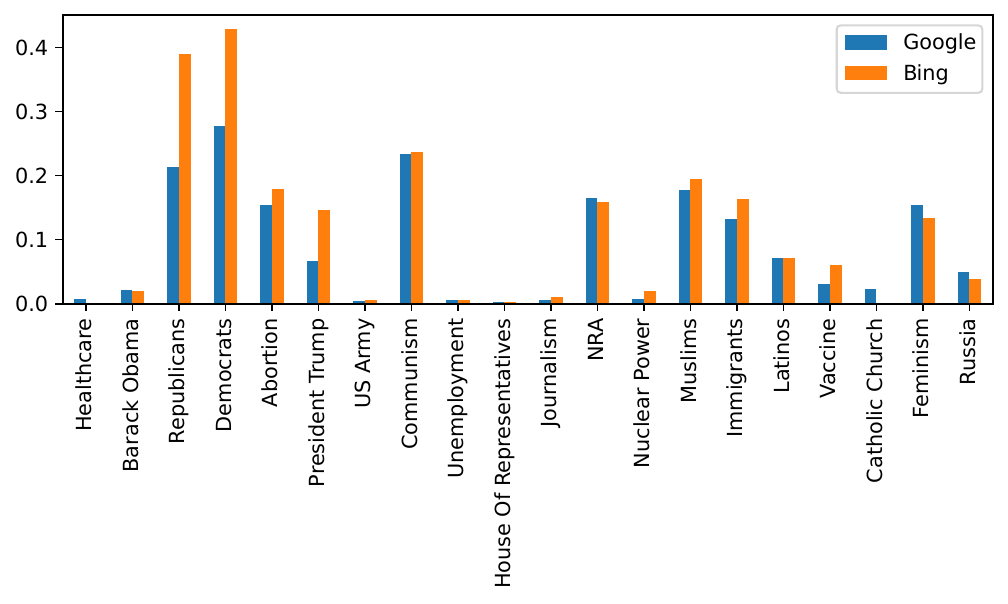} 
    
    \caption{Average bias scores (calculated on biased and unbiased query suggestions) per category for both search engines.}
    \label{fig:comparison_bars}
\end{figure}

\textbf{Dataset.}
We use part of the \texttt{$Q_{bias}$} dataset~\cite{Haak23_qbias}. 
\texttt{$Q_{bias}$} contains numerous search query suggestions collected from Google and Bing for topics from AllSides balanced news~\cite{allsides2}. 
Suggestions were collected for the root query, which was extended by the letters \texttt{a} to \texttt{z}, e.g., ``democrats a'', ``democrats b'', and so on.
As our slice, we select (and deduplicate) all search query suggestions for 20 topics from \texttt{$Q_{bias}$} (for topics see \autoref{fig:comparison_bars}), that have a nearly full set of 270 (10 for the base query, 10 each for a-z extensions) suggestions for Bing and Google, indicating a high search volume and a high interest in these topics.
Note that Google has considerably fewer suggestions, possibly due to their filtering of biased suggestions.
\autoref{tab:numbers} describes the size of our used dataset at different filter stages.

\begin{table}[b]
\small
\caption{Number of query suggestions at our filter stages.}
\begin{tabular}{l|cccc}
       & Slice of \texttt{$Q_{bias}$} & Deduplicated & Biased & \% Biased \\ \hline
Bing   & 5,380       & 5,009                & 1,051            & 21               \\
Google & 4,295       & 4,048                & 678             & 16.7               \\
\hline
Total  & 9,675       & 9,057                & 1,729            & 19.1           
\end{tabular}
\label{tab:numbers}
\end{table}

\textbf{Bias Identification.}
To identify potentially biased search query suggestions and evaluate GPT's ability to quantify bias, we provide GPT with the text along with a prompt 
defining bias\footnote{We give the following guidelines: \textit{``Does it carry political ideology bias, or contain potentially harmful, toxic or insulting words, false information or connotation? Could it offend someone or produce search results that could propagate a biased ideology or certain political agenda? Is it okay to be read by a child?''}} and asking to rate texts on a scale from 0 (not biased) to 100 (very biased).
Our Prompt is based on a general concept of biased language as non-neutral~\cite{Hube_2018b} and opinionated~\cite{Recasens_2013}. 
We incorporated aspects of partisan bias~\cite{Gawronski_partisan}, (political) stance~\cite{liu2021}, ideological bias~\cite{Mokhberian_2020}, linguistic bias~\cite{chen-etal-2020-analyzing}, and related concepts such as toxicity and offensive or profane language.
By utilizing a rather vague formulation, especially for the latter aspect, we try to let the model's concept of bias (severity) for query suggestions be the guideline for the judgment. 
GPT assigned scores in five distinct ranges (81\% at 0-10, 6\% at 20, 5\% at 30, 5\% at 50, and 3\% at 65-85). 
This indicates the model's indecisiveness or inability to assign a fine-grained absolute score.
Manual examination showed 0 or 10 being assigned to unbiased texts, so we consider texts with scores greater than 10 as biased.
Based on our manual evaluation, we find the differentiation of biased and not biased to be successful.

\textbf{Bias Quantification.}
We quantify the texts' bias intensity by producing a score based on many pairwise relative bias assessments combined with applying an Elo algorithm~\cite{good_Elo}.
To determine the absolute bias of texts, a score must be in an appropriate relationship to other scores.
For this reason, we use relative judgment assessments to construct the scores.
The number of pairs is determined by a simulation that indicates that after eight rounds of comparisons, the rank correlation between the produced ranking and an ideal ranking of scores is 0.85, with diminishing returns from more rounds of comparisons. 
That means that for our dataset of 1,729 biased texts, 6,916 comparisons are made.
Using the same definition of bias as in the bias identification step, GPT determines the more biased query for each pair.
Based on these pairwise relative judgments, we derive a 0 to 1 scaled bias score using an Elo algorithm from the chess domain.
The assumption when determining a (chess) player's score is that the stronger player is likely to win in a match.
However, a player can still lose to a weaker player.
After an increasing number of matches, the Elo rating converges with the actual playing strength~\cite{boubdir2023elo}, while in our case, this would be the actual bias intensity of text, respectively.
The approach fits our needs because different users might perceive the bias intensity differently, similar to how a match can go one way or another. 
After a given set of comparisons and applying the Elo algorithm, we obtain scores representing the order of the bias intensity in a pool of biased queries.

If $T_1$ and $T_2$ have identical ratings, the estimated probability of $T_1$ winning is 50\%.
Initially, Elo scores of all texts are given a predefined value of 1200.
The expected probability $E_{T_1}$ that $T_1$ with Elo rating $R_1$ wins against $T_2$ with Elo rating $R_2$ is defined as follows\cite{good_Elo, boubdir2023elo}:

\begin{equation}
    E_{T_1} = \frac{1}{1 + 10^{(R_2-R_1)/400}}.
\end{equation}
After a victory of $T_1$ over $T_2$, the new rating $R^{\prime}_{1}$ of $T_1$ is calculated as follows:
\begin{equation}
    R^{\prime}_{1} = R_{1} + k (S_{T_1}- E_{T_1}) .
\end{equation}
The constant $k$ (set to 16) controls the strength of the score change after a game. 
$S_T$ is 1 if $T$ has won and 0 otherwise. The expected probability of winning for $T_2$ and the calculation of the score update are calculated analogously.

\begin{figure}[t]
    \centering
      \includegraphics[width=1\linewidth]{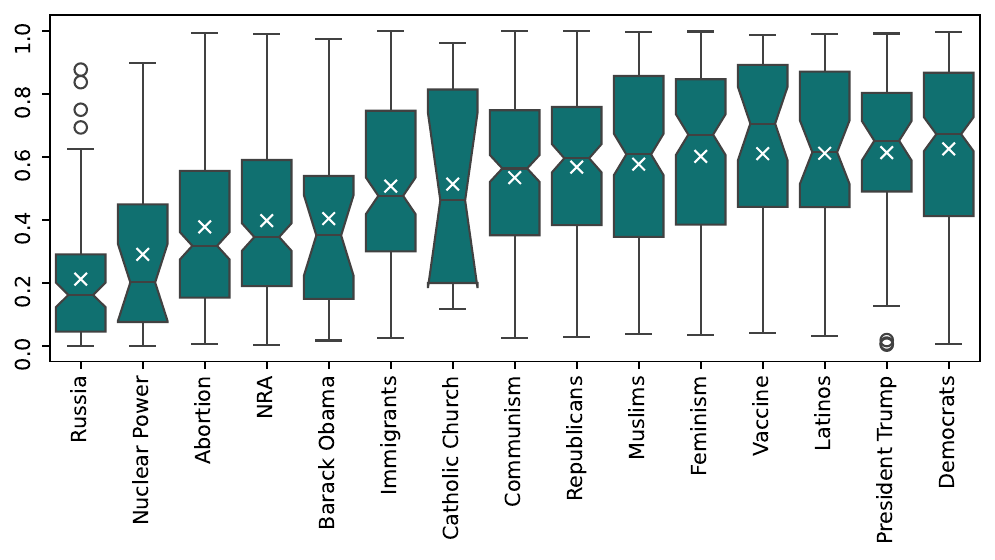}
    
    \caption{Boxplots indicating the distribution of bias scores per category (>~8 biased texts) ordered by mean ($x$ in plot). }
    \label{fig:comparison_boxplot}
\end{figure}

\section{Bias in Political News Search}\label{sec:bias}
We manually examine the ranking induced by the bias scores produced by our approach. 
Overall, we agree with the ranking. 
The ten most biased query suggestions contain ``feminism killed socialism'', ``democrats say men can get pregnant'', ``muslims dogs'', and ``feminism ruined movies''. 
The ten lowest scoring texts are much less biased, e.g., ``nuclear power bombs'', ``russia vs ucraina'', and ``abortion California''.
The values depicted in \autoref{fig:comparison_bars} reflect the average bias score per suggestion for the given categories.
For most topics, the average bias score is similar for both search engines. 
However, the topics ``Republicans'' and ``Democrats'' highlight a more substantial bias in Bing's suggestions.
As shown in \autoref{fig:comparison_boxplot}, all topics have a notable variance. 
We found the most severely biased suggestions on the topic ``Democrats''.
Overall, the intra-topic variance tends to be greater than the inter-topic variance, which might be enforced by selecting only relatively controversial topics.

\section{Limitations}
We have evaluated the ability of our approach to utilize GPT for bias quantification in a qualitative manner. 
However, we have conducted further analyses of the results produced by our approach for text simplicity, another subjective aspect of language. 
We compared the ranking produced by human annotators for the pairwise decision task with the ranking produced by GPT models for the same task, given the same prompt.
Here, we achieved high correlations and agreement between the rankings, with Cohen's~$\kappa$ of 0.66, Spearmans'~$\rho$ of 0.8, and Kendall's~$\tau$ of 0.62.
Although these results and those discussed in \autoref{sec:bias} indicate generally high suitability of the approach, direct comparison with human-annotated data would be beneficial.
A debatable aspect of our work is whether large language models should be utilized in performing socially normative evaluation. However, we, like others before, have shown that these models are generally capable of effectively fulfilling this task. 
Whether these models have intrinsic ideological biases is not part of our investigation and would have to be investigated further.

\section{Conclusion}
In this ongoing work, we show that large language models can be utilized to identify biased search queries and that pairwise comparison in conjunction with Elo-based ranking is capable of producing effective bias scores. 
We identified substantial biases in search query suggestions in the U.S. political news domain.

\bibliographystyle{ACM-Reference-Format}
\bibliography{lib}


\begin{thebibliography}{20}


\ifx \showCODEN    \undefined \def \showCODEN     #1{\unskip}     \fi
\ifx \showDOI      \undefined \def \showDOI       #1{#1}\fi
\ifx \showISBNx    \undefined \def \showISBNx     #1{\unskip}     \fi
\ifx \showISBNxiii \undefined \def \showISBNxiii  #1{\unskip}     \fi
\ifx \showISSN     \undefined \def \showISSN      #1{\unskip}     \fi
\ifx \showLCCN     \undefined \def \showLCCN      #1{\unskip}     \fi
\ifx \shownote     \undefined \def \shownote      #1{#1}          \fi
\ifx \showarticletitle \undefined \def \showarticletitle #1{#1}   \fi
\ifx \showURL      \undefined \def \showURL       {\relax}        \fi
\providecommand\bibfield[2]{#2}
\providecommand\bibinfo[2]{#2}
\providecommand\natexlab[1]{#1}
\providecommand\showeprint[2][]{arXiv:#2}

\bibitem[AllSides(2021)]%
        {allsides2}
\bibfield{author}{\bibinfo{person}{AllSides}.} \bibinfo{year}{2021}\natexlab{}.
\newblock \bibinfo{title}{How AllSides Creates Balanced News: A Step-by-Step Guide}.
\newblock
\newblock
\urldef\tempurl%
\url{allsides.com/blog/how-does-allsides-create-balanced-news}
\showURL{%
Retrieved Nov 30, 2022 from \tempurl}


\bibitem[Bonart et~al\mbox{.}(2019)]%
        {Bonart_2019}
\bibfield{author}{\bibinfo{person}{M. Bonart}, \bibinfo{person}{A. Samokhina}, \bibinfo{person}{G. Heisenberg}, {and} \bibinfo{person}{P. Schaer}.} \bibinfo{year}{2019}\natexlab{}.
\newblock \showarticletitle{{An investigation of biases in web search engine query suggestions}}.
\newblock \bibinfo{journal}{\emph{OIR}} \bibinfo{volume}{44}, \bibinfo{number}{2} (\bibinfo{year}{2019}), \bibinfo{pages}{365–381}.
\newblock
\showISSN{1468-4527}


\bibitem[Boubdir et~al\mbox{.}(2023)]%
        {boubdir2023elo}
\bibfield{author}{\bibinfo{person}{M. Boubdir}, \bibinfo{person}{E. Kim}, \bibinfo{person}{B. Ermis}, \bibinfo{person}{S. Hooker}, {and} \bibinfo{person}{M. Fadaee}.} \bibinfo{year}{2023}\natexlab{}.
\newblock \bibinfo{title}{Elo Uncovered: Robustness and Best Practices in Language Model Evaluation}.
\newblock
\newblock
\showeprint[arxiv]{2311.17295}~[cs.CL]


\bibitem[Chen et~al\mbox{.}(2020)]%
        {chen-etal-2020-analyzing}
\bibfield{author}{\bibinfo{person}{Wei-Fan Chen}, \bibinfo{person}{Khalid Al~Khatib}, \bibinfo{person}{Henning Wachsmuth}, {and} \bibinfo{person}{Benno Stein}.} \bibinfo{year}{2020}\natexlab{}.
\newblock \showarticletitle{Analyzing Political Bias and Unfairness in News Articles at Different Levels of Granularity}. In \bibinfo{booktitle}{\emph{NLPCSS}}. \bibinfo{pages}{149--154}.
\newblock
\urldef\tempurl%
\url{https://doi.org/10.18653/v1/2020.nlpcss-1.16}
\showDOI{\tempurl}


\bibitem[Epstein and Robertson(2015)]%
        {epstein2015}
\bibfield{author}{\bibinfo{person}{R. Epstein} {and} \bibinfo{person}{R.E. Robertson}.} \bibinfo{year}{2015}\natexlab{}.
\newblock \showarticletitle{The search engine manipulation effect ({SEME}) and its possible impact on the outcomes of elections}.
\newblock \bibinfo{journal}{\emph{PNAS}} \bibinfo{volume}{112}, \bibinfo{number}{33}, \bibinfo{pages}{E4512--E4521}.
\newblock
\showISSN{0027-8424, 1091-6490}


\bibitem[Gawronski(2021)]%
        {Gawronski_partisan}
\bibfield{author}{\bibinfo{person}{Bertram Gawronski}.} \bibinfo{year}{2021}\natexlab{}.
\newblock \showarticletitle{Partisan bias in the identification of fake news}.
\newblock \bibinfo{journal}{\emph{TiCS}} \bibinfo{volume}{25}, \bibinfo{number}{9} (\bibinfo{year}{2021}), \bibinfo{pages}{723--724}.
\newblock


\bibitem[Good(1955)]%
        {good_Elo}
\bibfield{author}{\bibinfo{person}{I.~J. Good}.} \bibinfo{year}{1955}\natexlab{}.
\newblock \showarticletitle{On the Marking of Chess-Players}.
\newblock \bibinfo{journal}{\emph{The Mathematical Gazette}} \bibinfo{volume}{39}, \bibinfo{number}{330} (\bibinfo{year}{1955}), \bibinfo{pages}{292--296}.
\newblock
\showISSN{00255572}
\urldef\tempurl%
\url{http://www.jstor.org/stable/3608567}
\showURL{%
\tempurl}


\bibitem[Haak and Schaer(2021)]%
        {haak_perception-aware_2021}
\bibfield{author}{\bibinfo{person}{F. Haak} {and} \bibinfo{person}{P. Schaer}.} \bibinfo{year}{2021}\natexlab{}.
\newblock \showarticletitle{Perception-{Aware} {Bias} {Detection} for {Query} {Suggestions}}. In \bibinfo{booktitle}{\emph{{BIAS}}}. \bibinfo{pages}{130--142}.
\newblock
\showISBNx{978-3-030-78818-6}


\bibitem[Haak and Schaer(2022)]%
        {HaakS22}
\bibfield{author}{\bibinfo{person}{F. Haak} {and} \bibinfo{person}{P. Schaer}.} \bibinfo{year}{2022}\natexlab{}.
\newblock \showarticletitle{Auditing Search Query Suggestion Bias Through Recursive Algorithm Interrogation}. In \bibinfo{booktitle}{\emph{WebSci '22}}. \bibinfo{pages}{219--227}.
\newblock


\bibitem[Haak and Schaer(2023)]%
        {Haak23_qbias}
\bibfield{author}{\bibinfo{person}{F. Haak} {and} \bibinfo{person}{P. Schaer}.} \bibinfo{year}{2023}\natexlab{}.
\newblock \showarticletitle{Qbias - A Dataset on Media Bias in Search Queries and Query Suggestions}. In \bibinfo{booktitle}{\emph{WebSci '23}}. \bibinfo{pages}{239–244}.
\newblock
\showISBNx{9798400700897}


\bibitem[Hube and Fetahu(2019)]%
        {Hube_2018b}
\bibfield{author}{\bibinfo{person}{Christoph Hube} {and} \bibinfo{person}{Besnik Fetahu}.} \bibinfo{year}{2019}\natexlab{}.
\newblock \showarticletitle{Neural Based Statement Classification for Biased Language}. In \bibinfo{booktitle}{\emph{WSDM}}. \bibinfo{publisher}{ACM}.
\newblock


\bibitem[Introna and Nissenbaum(2000)]%
        {Introna2000}
\bibfield{author}{\bibinfo{person}{L. Introna} {and} \bibinfo{person}{H. Nissenbaum}.} \bibinfo{year}{2000}\natexlab{}.
\newblock \showarticletitle{Defining the Web: the politics of search engines}.
\newblock \bibinfo{journal}{\emph{Computer}} \bibinfo{volume}{33}, \bibinfo{number}{1}, \bibinfo{pages}{54--62}.
\newblock


\bibitem[Kulshrestha et~al\mbox{.}(2019)]%
        {kulshrestha2019}
\bibfield{author}{\bibinfo{person}{J. Kulshrestha}, \bibinfo{person}{M. Eslami}, \bibinfo{person}{J. Messias}, \bibinfo{person}{M.~B. Zafar}, \bibinfo{person}{S. Ghosh}, \bibinfo{person}{K.~P. Gummadi}, {and} \bibinfo{person}{K. Karahalios}.} \bibinfo{year}{2019}\natexlab{}.
\newblock \showarticletitle{Search bias quantification: investigating political bias in social media and web search}.
\newblock \bibinfo{journal}{\emph{Inf. Retr. J.}} \bibinfo{volume}{22}, \bibinfo{number}{1}, \bibinfo{pages}{188--227}.
\newblock
\showISSN{1573-7659}


\bibitem[Liu et~al\mbox{.}(2021)]%
        {liu2021}
\bibfield{author}{\bibinfo{person}{Ruibo Liu}, \bibinfo{person}{Chenyan Jia}, {and} \bibinfo{person}{Soroush Vosoughi}.} \bibinfo{year}{2021}\natexlab{}.
\newblock \showarticletitle{A Transformer-based Framework for Neutralizing and Reversing the Political Polarity of News Articles}.
\newblock \bibinfo{journal}{\emph{Proc. ACM Hum.-Comput. Interact.}}  \bibinfo{volume}{5}, \bibinfo{pages}{1--26}.
\newblock
\showISSN{2573-0142}


\bibitem[Mokhberian et~al\mbox{.}(2020)]%
        {Mokhberian_2020}
\bibfield{author}{\bibinfo{person}{Negar Mokhberian}, \bibinfo{person}{Andr{\'{e} }s Abeliuk}, \bibinfo{person}{Patrick Cummings}, {and} \bibinfo{person}{Kristina Lerman}.} \bibinfo{year}{2020}\natexlab{}.
\newblock \showarticletitle{Moral Framing and Ideological Bias of News}.
\newblock In \bibinfo{booktitle}{\emph{SocInfo}}. \bibinfo{pages}{206--219}.
\newblock


\bibitem[Niu and Kelly(2014)]%
        {niu}
\bibfield{author}{\bibinfo{person}{Xi Niu} {and} \bibinfo{person}{Diane Kelly}.} \bibinfo{year}{2014}\natexlab{}.
\newblock \showarticletitle{The use of query suggestions during information search}.
\newblock \bibinfo{journal}{\emph{IPM}} \bibinfo{volume}{50}, \bibinfo{number}{1}, \bibinfo{pages}{218--234}.
\newblock
\showISSN{0306-4573}


\bibitem[Ray(2020)]%
        {ray2020}
\bibfield{author}{\bibinfo{person}{L. Ray}.} \bibinfo{year}{2020}\natexlab{}.
\newblock \bibinfo{title}{2020 Google Search Survey: How Much Do Users Trust Their Search Results?}
\newblock
\newblock
\urldef\tempurl%
\url{moz.com/blog/2020-google-search-survey}
\showURL{%
Retrieved Nov 30, 2022 from \tempurl}


\bibitem[Recasens et~al\mbox{.}(2013)]%
        {Recasens_2013}
\bibfield{author}{\bibinfo{person}{Marta Recasens}, \bibinfo{person}{Cristian Danescu-Niculescu-Mizil}, {and} \bibinfo{person}{Dan Jurafsky}.} \bibinfo{year}{2013}\natexlab{}.
\newblock \showarticletitle{Linguistic Models for Analyzing and Detecting Biased Language}. In \bibinfo{booktitle}{\emph{ACL}}. \bibinfo{pages}{1650--1659}.
\newblock


\bibitem[Robertson et~al\mbox{.}(2018)]%
        {robertson_auditing_2018}
\bibfield{author}{\bibinfo{person}{R.~E. Robertson}, \bibinfo{person}{S. Jiang}, \bibinfo{person}{K. Joseph}, \bibinfo{person}{L. Friedland}, \bibinfo{person}{D. Lazer}, {and} \bibinfo{person}{C. Wilson}.} \bibinfo{year}{2018}\natexlab{}.
\newblock \showarticletitle{Auditing Partisan Audience Bias within Google Search}.
\newblock \bibinfo{journal}{\emph{Proc. ACM Hum.-Comput. Interact.}} \bibinfo{volume}{2}, \bibinfo{number}{CSCW}, Article \bibinfo{articleno}{148} (\bibinfo{year}{2018}).
\newblock


\bibitem[Spinde et~al\mbox{.}(2023)]%
        {Spinde2023MediaBias}
\bibfield{author}{\bibinfo{person}{T. Spinde}, \bibinfo{person}{S. Hinterreiter}, \bibinfo{person}{F. Haak}, \bibinfo{person}{T. Ruas}, \bibinfo{person}{H. Giese}, \bibinfo{person}{N. Meuschke}, {and} \bibinfo{person}{B. Gipp}.} \bibinfo{year}{2023}\natexlab{}.
\newblock \showarticletitle{The Media Bias Taxonomy: A Systematic Literature Review on the Forms and Automated Detection of Media Bias}.
\newblock  (\bibinfo{year}{2023}).
\newblock
\showeprint[arxiv]{2312.16148}~[cs.CL]


\end{thebibliography}

\end{document}